\documentclass[conference]{IEEEtran}
\IEEEoverridecommandlockouts
\usepackage{orcidlink}

\usepackage{textcomp}
\usepackage{xcolor}

\usepackage{cite}

\usepackage{siunitx}

\usepackage{graphicx}
\usepackage{caption}
\usepackage{subcaption} 
\graphicspath{{./figures/}}  
\DeclareGraphicsExtensions{.pdf,.jpeg,.png}

\usepackage{amsmath, amssymb, amsfonts}
\usepackage{amsthm}

\usepackage{enumitem}

\usepackage{array}

\usepackage{optidef}

\usepackage{algorithm}                 
\usepackage[noend]{algpseudocode}      

\algrenewcommand\algorithmicrequire{\textbf{Input:}}
\algrenewcommand\algorithmicensure{\textbf{Output:}}

\algtext*{EndWhile} 
\algtext*{EndIf}   
\algtext*{EndFunction}

\usepackage{url}
\usepackage{booktabs} 

\usepackage{cleveref} 

\def\BibTeX{{\rm B\kern-.05em{\sc i\kern-.025em b}\kern-.08em
    T\kern-.1667em\lower.7ex\hbox{E}\kern-.125emX}}

\newtheorem{thm}{\bf Theorem}

\begin{document}
\bstctlcite{IEEEexample:BSTcontrol}


\title{FlexSAN: A Flexible Regenerative Satellite Access Network Architecture} 

\author{
  \IEEEauthorblockN{
    Weize Kong\IEEEauthorrefmark{1},
    Chaoqun You\IEEEauthorrefmark{1},
    Xuming Pei\IEEEauthorrefmark{1},
    YueGao\IEEEauthorrefmark{1}
  }
  \IEEEauthorblockA{\IEEEauthorrefmark{1}Fudan University}
}

\maketitle

\begin{abstract}
The regenerative satellite access network (SAN) architecture deploys next-generation NodeB (gNBs) on satellites to enable enhanced network management capabilities. It supports two types of regenerative payload, on-board gNB and on-board gNB-Distributed Unit (gNB-DU). Measurement results based on our prototype implementation show that the on-board gNB offers lower latency, while the on-board gNB-DU is more cost-effective, and there is often a trade-off between Quality-of-Service (QoS) and operational expenditure (OPEX) when choosing between the two payload types. However, current SAN configurations are \textit{static} and \textit{inflexible} --- either deploying the full on-board gNB or only the on-board gNB-DU. This rigidity can lead to resource waste or poor user experiences. In this paper, we propose Flexible SAN (FlexSAN), an adaptive satellite access network architecture that dynamically configures the optimal regenerative payload based on real-time user demands. FlexSAN selects the lowest OPEX payload configuration when all user demands are satisfied, and otherwise maximizes the number of admitted users while ensuring QoS for connected users. To address the computational complexity of dynamic payload selection, we design an adaptive greedy heuristic algorithm. Extensive experiments validate FlexSAN's effectiveness, showing a 36.1\% average improvement in user admission rates and a 15\% OPEX reduction over static SANs.
\end{abstract}

\begin{IEEEkeywords}
SAN, 
Regenerative Payload, 
Flexibility
\end{IEEEkeywords}

\section{Introduction}
To realize the ambitious vision of the "Internet of Everything" (IoE) in future Sixth-Generation (6G) wireless networks, Non-Terrestrial Networks (NTNs) have emerged as essential enablers of global seamless connectivity \cite{10500741, 9217520, 9406391}. Central to NTNs is the Satellite Access Network (SAN), which incorporates aerial and space nodes such as Unmanned Aerial Vehicles (UAVs), High Altitude Platforms (HAPs), and satellites to complement terrestrial networks, bridging coverage gaps and enabling high-speed broadband services \cite{9970355,10978515,9385374,3gpp.38.821}. SANs significantly empower Mobile Network Operators (MNOs) to expand service areas efficiently, ensuring ubiquitous connectivity and economic sustainability \cite{10572042}.

\begin{figure}[t!]
  \centering
  \subfloat[On-board gNB]{\includegraphics[width=0.8\columnwidth]{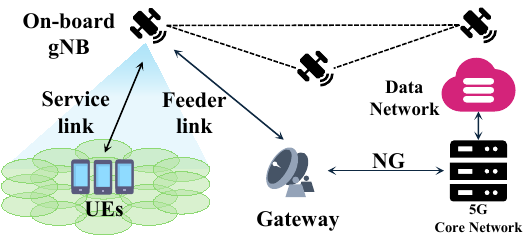}} \\
  \subfloat[On-board gNB-DU]{\includegraphics[width=0.8\columnwidth]{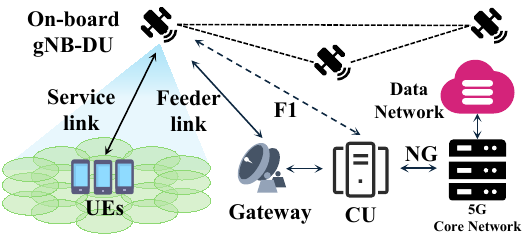}}
  \caption{Regenerative payload architecture.}
  \label{fig:vertical_subfigs}
\end{figure}

Current 3GPP standards outline two primary SAN architectures: the transparent payload, which functions as a simple radio repeater without any signal processing capability; and the regenerative payload, which incorporates advanced signal processing directly on the satellite itself, allowing the satellite to decode, process, and retransmit signals independently of ground stations \cite{3gpp.23.737,3gpp.38.801,9722775}. 
Our work focuses on regenerative architecture due to its enhanced network management capabilities and support for advanced 5G/6G features, which represent the future development direction of 6G. 
As illustrated in Fig.~\ref{fig:vertical_subfigs}, the regenerative architecture offers two types of payloads, (i) on-board gNB (shown in Fig.~\ref{fig:vertical_subfigs}(a)), where the complete gNB is deployed on the satellite; and (ii) on-board gNB-DU (shown in Fig.~\ref{fig:vertical_subfigs}(b)), where the traditional monolithic gNB is split into a Distributed Unit (DU) and a Centralized Unit (CU), with only the DU placed on the satellite while the CU functions are offloaded to ground stations \cite{9632811,10.1145/3686138.3686140}. We implemented a SAN prototype using OpenAirInterface (OAI) \cite{oai}, and the measurement results show that the on-board gNB architecture offers lower service latency, while the on-board gNB-DU architecture consumes fewer onboard resources. Therefore, there is always a trade-off between user quality of service (QoS) and operational expenditure (OPEX) when choosing between the two regenerative SAN architectures.

Current SANs often rely on \textit{static} payload configurations --- either on-board gNB or on-board gNB-DU. However, a core characteristic of SAN is its dynamic nature. The high-speed movement of satellites, link quality fluctuations, and the constantly changing user equipment (UE) demands lead to frequent changes in network conditions and rapid fluctuations in the QoS UE experiences. Using static payload configurations in a highly dynamic SAN often results in either unmet user demands or wasted SAN resources~\cite{9107209}. Therefore, architectural configurations in SANs need to be \textit{flexible}, enabling the SAN architecture to adapt dynamically to changing conditions.

Despite the extensive research on radio access network (RAN) architecture design in terrestrial networks \cite{8877874, 10.1145/3452296.3472894, flex5g}, only a few studies consider SAN. \cite{10901285} proposed an energy-efficient SAN architecture. Nevertheless, its primary goal is to minimize energy consumption instead of flexibility. Meanwhile, \cite{campana2023ran,10479650,10597025, 9781611, d2022orchestran} have highlighted the urgent necessity for a flexible and dynamic SAN architecture. However, these works have neither quantified the advantages and disadvantages of the two architectures nor addressed the issue of flexible access in the SAN architecture. Designing an adaptive SAN architecture that dynamically configures the optimal regenerative payload, however, is by no means an easy problem. This is because (i) the impact of UE scale on each UE's access architecture configuration is \textit{unclear}. With a small UE scale, the satellite has ample resources that users can afford to waste; but with a large UE scale, on-board resources become limited, and UEs must use them carefully and efficiently. (ii) Current solutions often require \textit{long computation times}, which cannot keep up with the highly dynamic nature of SANs --- by the time a result is obtained, the state of the SAN may have already changed.

In this paper, to address the above challenges, we propose FlexSAN that provides flexible regenerative payloads in SAN for UEs to dynamically adapt to. This is done by quantifying and analysing two different user load scenarios. When the number of UEs is low, the optimization objective is to minimize on-board resource consumption while satisfying the QoS requirements of all users. When the number of UEs is high, the objective shifts to maximizing the number of admitted users while ensuring that the QoS requirements of all admitted users are met. Solving the above two optimization problems remains challenging, as traditional methods for computing the optimal solution are often of high complexity, resulting in long computation times. To enable solutions within polynomial time, we design two greedy algorithms to address these two problems.

The contributions are summarized as follows,
\begin{itemize}
    \item We are the first to quantify the trade-off between the two regenerative SAN payload types. We first establish an implementation platform to measure the computational cost and the latency of both payload types, and then use mathematical language to describe across these dimensions. (Section~\ref{motivation})
    \item We propose FlexSAN, a flexible regenerative SAN architecture. FlexSAN contains two stages. When the UE scale is small, it is desired to minimize the computation resource usage on-board; and when the UE scale is large, it is desired to maximize the number of accepted UEs to get services. (Section~\ref{problem})
    \item We design greedy algorithms for the above two optimization problems. The solutions are capable of delivering near-optimal solutions in polynomial time, suitable for real-time deployment. (Section~\ref{algorithm})   
    \item We extensively evaluate FlexSAN through simulations, demonstrating its significant advantages over conventional static architectures in reducing operational expenses and enhancing user admission rates under dynamic network conditions. (Section~\ref{performance})
\end{itemize}

\section{Backgrounds, Motivation and Challenges} \label{motivation}

\subsection{Backgrounds} \label{sec:backgrounds}

\noindent\textbf{Functional Split}: Functional split is a principle central to network virtualization that enables flexible deployments by dividing the gNB protocol stack into gNB-CU and gNB-DU. While this paradigm is well-established in terrestrial networks \cite{10901285, flex5g}, its adoption in Satellite Access Networks (SANs) is critically constrained by the long propagation delays inherent in satellite links.

\begin{figure}[t!]
\centering 
\includegraphics[width=\columnwidth]{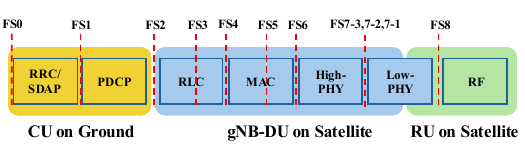}
\caption{Illustration of 3GPP RAN functional split options (FS0--FS8) in the gNB protocol stack for SANs \cite{larsen2018survey}.}
\label{fig:split_options}
\end{figure}

As illustrated in Fig.~\ref{fig:split_options}, the 3GPP standards define multiple split options. However, lower-layer splits (Options 3-8) demand sub-millisecond fronthaul latencies, a requirement that is unachievable over satellite feeder links \cite{campana2023ran}. This fundamental limitation effectively narrows the practical choices for SANs to two high-level architectures:

\begin{itemize}
\item \textbf{On-board gNB (Split 0):} All gNB functions, encompassing both the gNB-CU and the gNB-DU, are executed on the satellite. This configuration minimizes latency but imposes significant on-board computational and power demands.
\item \textbf{On-board gNB-DU (Split 2):} Only the lower-layer gNB-DU functions (RLC, MAC, PHY) reside on the satellite, while the computation-intensive gNB-CU functions (PDCP, RRC) are offloaded to a ground station. This split is increasingly favored for NTNs as it reduces payload complexity \cite{10572042, campana2023ran}.
\end{itemize}

\vspace{3pt}
\noindent\textbf{F1 interface}: In the Split 2 architecture, the on-board gNB-DU and the ground-based gNB-CU are logically connected via the F1 interface. A critical consequence in SANs is that all F1 signaling must traverse the satellite feeder link. Compared to terrestrial fiber connections, satellite feeder links are characterized by significantly longer propagation delays, lower capacity, and reduced reliability. As a result, the F1 interface signaling between the on-board gNB-DU and the ground-based gNB-CU introduces significant additional latency, often exceeding 80\,ms per round trip \cite{10572042, campana2023ran}. Despite this latency challenge, the F1 interface is fully standardized by 3GPP, which ensures interoperability and makes it a well-defined and practical option for satellite-based deployments \cite{3gpp.38.470, 3gpp.38.471, 3gpp.38.472, 3gpp.38.473, 3gpp.38.474, 3gpp.38.475}.

\subsection{Motivations}

We designed and implemented a SAN prototype using the OAI platform, as depicted in Fig.~\ref{fig:prototype}(a). Critically, our prototype operates directly on the 3GPP-specified NTN band (Band 256) at a 2200\,MHz carrier frequency and 10\,MHz bandwidth, emulating a realistic low earth orbit (LEO) satellite at a 359.7\,km altitude. This approach ensures greater real-world relevance than terrestrial emulations. The built prototype operates on a commercial laptop equipped with a 2.7\,GHz CPU core and 24\,GB RAM, offering a practical yet capable compute environment for testing. Using Linux utilities such as \texttt{tc} \cite{linux_tc_man} for propagation delay injection and \texttt{pidstat} \cite{sysstat} for CPU monitoring, we conducted a comparative evaluation of the two primary static architectures: the on-board gNB and the on-board gNB-DU.

\begin{figure}[t!]
\centering
\includegraphics[width=0.49\textwidth]{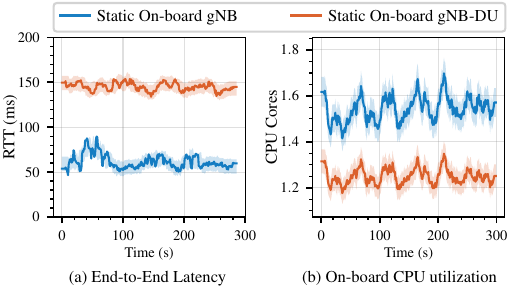}
\caption{Experimental results: (a) End-to-end latency, and (b) CPU utilization on the on-board gNB and on-board gNB-DU.}
\label{fig:sim_results}
\end{figure}

Our experiments first confirmed a clear and quantifiable architectural dilemma, as shown in Fig.~\ref{fig:sim_results}. The on-board gNB architecture consistently delivered low end-to-end latency (61.4\,ms), ideal for delay-sensitive services. In contrast, the on-board gNB-DU architecture, while reducing on-board CPU load by nearly 20\%, incurred a prohibitively high deterministic latency (145.1\,ms) due to F1 interface signaling exchange over the feeder link. This exposes a rigid, system-wide trade-off: operators must choose between significant OPEX savings and the ability to deliver high-value, low-latency services.

More critically, we found that this static trade-off masks a deeper, dynamic inefficiency. By introducing varying traffic loads with \texttt{iperf3} \cite{iperf3}, we observed that the average Round-Trip Time (RTT) per user escalates sharply with the number of concurrent users, irrespective of their individual data rates (Fig.~\ref{fig:prototype}(b)). The RTT surged from 60-120\,ms with a single UE to over 500\,ms with six active UEs. This demonstrates that the aggregate computational load and resulting queuing delays become the dominant factor in user-experienced latency. Consequently, even users on the supposedly low-latency on-board gNB architecture suffer from severe QoS degradation under high load.

These findings lead to a crucial \textit{insight} that static SAN architectures are fundamentally incapable of handling dynamic network conditions. They are trapped by a rigid trade-off at low loads and suffer from systemic performance degradation at high loads. This motivates the need for a new paradigm --- a flexible, adaptive architecture that can dynamically orchestrate resources and functional splits on a per-user basis to break this bottleneck. This is the core motivation for FlexSAN.

\begin{figure}[t!]
    \centering
    \begin{subfigure}{0.48\linewidth}
        \centering
        \includegraphics[width=\linewidth]{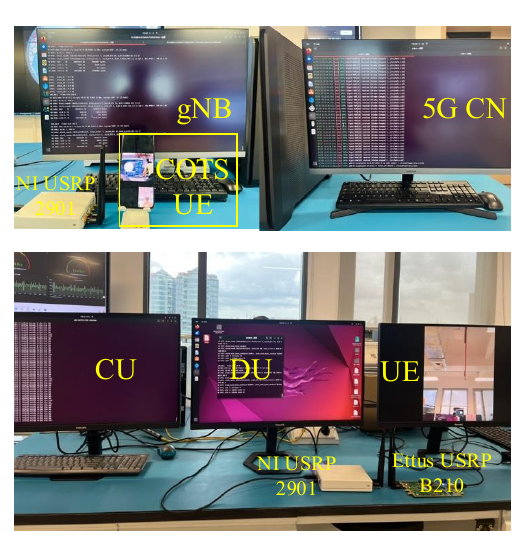}
        \caption{OAI-based SAN Prototype}
        \label{fig:prototype}
    \end{subfigure}
    \begin{subfigure}{0.48\linewidth}
        \centering
        \includegraphics[width=\linewidth]{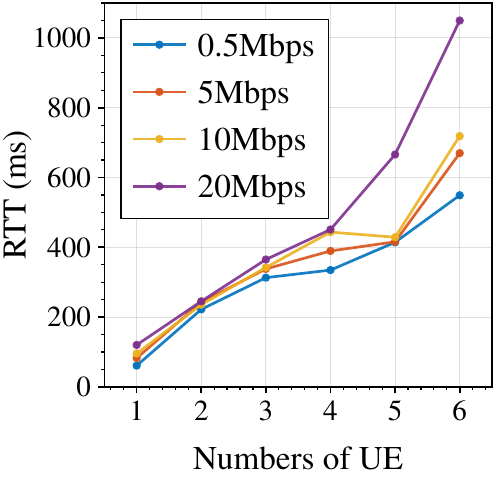}
        \caption{RTT Escalation under Load}
        \label{fig:rtt_vs_ue}
    \end{subfigure}
    \vspace{4pt}
    \caption{The OAI-based SAN prototype and RTT measurements under varying load conditions.}
    \label{fig:prototype}
\end{figure}

\subsection{Challenges}

The design of flexible and dynamic SAN is fundamentally constrained by performance interdependence among users sharing onboard computational resources. Individual delays are influenced not only by personal architectural choices but also by collective resource usage and queuing effects. Achieving optimal system performance requires a holistic approach that jointly considers resource allocation, admission control, and regenerative payload selection. The intricate interplay among these elements significantly affects user QoS, yet accurately modeling their relationships remains a major challenge. Moreover, practical solutions must maintain extremely low computational complexity to support real-time operation as the user scale varies. Only under these conditions can such strategies be effectively applied in dynamic SAN environments.

\section{System Model} \label{system}

\subsection{Network Model}
\begin{figure}[t!]
  \centering
  \includegraphics[width=0.85\columnwidth]{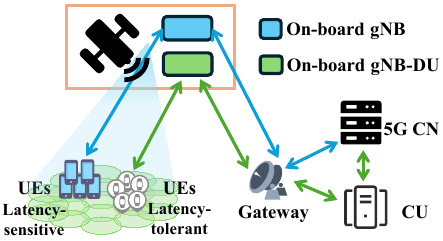} 
  \caption{Overview of the FlexSAN architecture.}
  \label{fig:overview}
\end{figure}

We consider a SAN where multiple users from the user set $\mathcal{U}$ require services from a satellite that covers them over discrete time slots $t \in \mathcal{T}$, as shown in Fig.~\ref{fig:overview}. There are two types of regenerative payloads for UEs to choose, on-board gNB $g=0$ and on-board gNB-DU $g=1$. Each UE $u \in \mathcal{U}$ specifies its maximum tolerable delay $T_u^{\text{max}}$ and minimum throughput $R_u^{\text{min}}$. The whole system is assumed to be virtualized to allow dynamic, per-user instantiation of two regenerative on-board payload types. 

At each time slot $t$, for UE $u\in\mathcal{U}$, we consider the decision variables to be as follows, 
\begin{itemize}
    \item $\mathbf{x}=[x_u^g(t)]$, defined as the regenerative architecture selection vector, where $x_u^g(t)=1$ if user $u$ is assigned to regenerative architecture $g \in \{0, 1\}$, where $g=0$ denotes on-board gNB and $g=1$ denotes on-board gNB-DU;
    \item $\mathbf{y} = [y_u(t)]$, defined as the UE admission vector, where the generic element $y_u(t)$ is a binary variable indicating whether UE $u$ is admitted by the SAN for task processing.
    \item $\mathbf{w}=[w_u(t)]$, defined as the bandwidth allocation vector, where the generic element $w_u(t)$ indicates the bandwidth allocated to UE $u$ at time slot $t$. $w_u(t) \ge 0$, and the total bandwidth available is $B_s$.
\end{itemize}

At this point, we are able to obtain the following constraint, that the sum of the selection indicators over all available regenerative payload types must equal the admission decision for each user. 
\begin{equation}
    \sum_{g \in \{0,1\}} x_u^g(t) = y_u(t), \quad \forall u \in \mathcal{U}
\end{equation}

This constraint ensures that a user can only be assigned to a regenerative payload if it is admitted, and that each admitted user is associated with exactly one regenerative architecture.

\subsection{Payload Architecture and Computational Cost}
The onboard computational load is a direct proxy for the operator's OPEX. Following the methodology in \cite{6771075, 9107209}, we model this cost based on the processing functions ($PF_i$, discussed in Section~\ref{motivation}) executed for each admitted user, measured in Giga Operations Per Second (GOPS).

The cost for user $u$, $C_g(u, t, w_u(t))$, depends on the assigned regenerative architecture $g$ and the allocated bandwidth $w_u(t)$. Let $\mathcal{P}_{\text{all}}$ be the complete set of processing functions and $\mathcal{P}_{\text{sat}} \subset \mathcal{P}_{\text{all}}$ be the subset remaining on the satellite for the gNB-DU architecture. The on-board computational cost generated by user $u$ is thus:
\begin{itemize}
    \item \textbf{On-board gNB ($g=0$):}
    \begin{equation}
        C_0(u, t, w_u(t)) = \sum_{i \in \mathcal{P}_{\text{all}}} c_i(u, t, w_u(t))
    \end{equation}
    \item \textbf{On-board gNB-DU ($g=1$):}
    \begin{equation}
        C_1(u, t, w_u(t)) = \sum_{i \in \mathcal{P}_{\text{sat}}} c_i(u, t, w_u(t))
    \end{equation}
\end{itemize}
The total on-board computational load is the sum of costs from all admitted users and is constrained by the satellite's total processing capacity, $C_{\text{Sat}}^{\text{Cap}}$:
\begin{equation}
\sum_{u \in \mathcal{U}} y_u(t) \sum_{g \in \{0, 1\}} x_u^g(t) C_g(u, t, w_u(t)) \le C_{\text{Sat}}^{\text{Cap}}
\end{equation}

\subsection{Transmission and Processing Delay}
The end-to-end delay serves as a critical determinant of user QoS. For an admitted user ($y_u(t)=1$), this total delay must not exceed the specified threshold $T_u^{\text{max}}$. This delay is composed of transmission and processing components.

\subsubsection{Transmission Delay}

The transmission delay on the service link, $t_{\text{trans},u}(t)$, is the sum of propagation delay and data transfer time. It is given by $d_u^s(t)/c + S_p / R_u^s(t, w_u(t))$, where $d_u^s(t)$ is the instantaneous distance to the satellite, $c$ is the speed of light, $S_p$ is the packet size, and $R_u^s(t, w_u(t))$ is the achievable rate. This rate depends on the allocated bandwidth $w_u(t)$ and the time-varying SNR $\xi_u^s(t)$, modeled as $R_u^s(t, w_u(t)) = w_u(t)\log_2(1+\xi_u^s(t))$. The throughput requirement is enforced as:
\begin{equation}
y_u(t)\cdot R_u^s(t, w_u(t)) \ge R_u^{\text{min}}
\end{equation}

While our experiments account for feeder link variations in Section~\ref{performance}, we omit their detailed modeling here for brevity.

\subsubsection{Processing Delay}
Processing delay, $t_u^{\text{proc}}(t)$, consists of a shared queuing delay and an architecture-specific F1 interface delay. We model the on-board processing unit as an M/M/1 queue, where the service rate is the satellite's capacity, $\mu_{\text{Sat}} = C_{\text{Sat}}^{\text{Cap}}$. The total arrival rate, $\lambda_{\text{Sat}}(t)$, is the aggregate on-board computation load from all admitted users (indexed by $u'$):
\begin{equation}
\lambda_{\text{Sat}}(t) = \sum_{u' \in \mathcal{U}} \sum_{g' \in \{0,1\}} x_{u'}^{g'}(t) C_{g'}(u', t, w_{u'}(t))
\end{equation}

The resulting queuing delay, which grows non-linearly with load, is $t_{\text{queue}}(t) = 1/(\mu_{\text{Sat}} - \lambda_{\text{Sat}}(t))$. The total processing delay for user $u$ also includes the F1 interface delay $t_{\text{F1}}(t)$ if the on-board gNB-DU architecture is selected:
\begin{equation}
t_u^{\text{proc}}(t) = t_{\text{queue}}(t) + x_u^1(t) \cdot t_{\text{F1}}(t)
\end{equation}

The total end-to-end latency for user $u$, $t_u(t, \mathbf{y}(t), \mathbf{x}(t), \mathbf{w}(t))$, is the sum of transmission and processing delays and must satisfy the user's QoS constraint:
\begin{equation}
    y_u(t) \cdot t_u(t, \mathbf{y}(t), \mathbf{x}(t), \mathbf{w}(t)) \le T_u^{\text{max}}
\end{equation}

\vspace{1pt}
\section{Problem Formulation} \label{problem}

Previous analyses reveal two fundamental limitations inherent in static SAN architectures. First, under normal network conditions, they impose a rigid trade-off between OPEX and QoS, forcing SANs to choose exclusively between low latency (on-board gNB) or low cost (on-board gNB-DU). Second, during network congestion, static SANs are unable to adapt to increased demands, resulting in severe performance degradation and an inability to meet users' service requirements.

To address these limitations, we formulate FlexSAN orchestration as a dynamic, two-stage optimization problem: 

\begin{itemize}
\item \textbf{Stage 1: OPEX Minimization (Normal Load)}. When resources are sufficient, FlexSAN dynamically orchestrates the regenerative payload to minimize computational cost (OPEX), while strictly meeting service requirements for all users. Unlike static designs, this allows simultaneous optimization of cost and QoS through per-user architectural and bandwidth adaptations.
\item \textbf{Stage 2: Service Maximization (Congestion)}. Under congestion, FlexSAN prioritizes maximizing user admissions, dynamically selecting which users to serve and how resources should be allocated to strictly satisfy their service demands. This adaptive approach ensures graceful performance degradation and robust QoS guarantees, significantly improving resilience compared to static configurations.
\end{itemize}

Below, we formally present the two optimization problems defining this adaptive approach.

\subsection{Problem $P_1$: OPEX Minimization under Normal Load}
Under normal operational conditions, when it is feasible to admit all users, the FlexSAN orchestrator's primary goal is to minimize the computational costs of the satellite payload. This optimization ensures cost-effective service delivery, defined formally as:
\begin{IEEEeqnarray}{cl}
\underset{
  \substack{
    \mathbf{x}, \mathbf{w}
  }
}{\text{min}} \quad &
  \displaystyle\sum_{u\in\mathcal{U}} \sum_{g\in\mathcal{G}}
  x_u^g(t) C_g\bigl(u, t, w_u(t)\bigr) \\[0.5ex]
\textit{s.t.}\quad
& t_u\bigl(t, \mathbf{x}(t), \mathbf{w}(t)\bigr) \le T_u^{\text{max}},\quad \forall u \in \mathcal{U} \label{c:p1_delay} \\[0.6ex]
& R_u^s\bigl(t, w_u(t)\bigr) \ge R_u^{\text{min}}, \hspace{1.6em} \quad \forall u \in \mathcal{U} \label{c:p1_throughput} \\[0.6ex]
& \sum_{g\in\mathcal{G}} x_u^g(t) = 1, \hspace{4.6em} \quad \forall u \in \mathcal{U} \label{c:p1_association} \\[0.6ex]
& \sum_{u\in\mathcal{U}} w_u(t) \le B_s \label{c:p1_service_bw} \\[0.6ex]
& \sum_{u\in\mathcal{U}} \sum_{g\in\mathcal{G}} x_u^g(t) C_g\bigl(u, t, w_u(t)\bigr) \le C_{\text{Sat}}^{\text{Cap}} \label{c:p1_sat_cpu}
\end{IEEEeqnarray}

Here, FlexSAN explicitly considers per-user service requirements (constraints~\eqref{c:p1_delay}, \eqref{c:p1_throughput}), total available bandwidth~\eqref{c:p1_service_bw}, and satellite computational capacity~\eqref{c:p1_sat_cpu}. By dynamically selecting architectures on a per-user basis (constraint~\eqref{c:p1_association}), FlexSAN achieves flexible trade-offs that static configurations inherently cannot.

\subsection{Problem $P_2$: Service Maximization under Congestion}
When the available network resources become insufficient to meet the demands of all users, FlexSAN dynamically transitions into a congestion-aware mode. Here, the orchestrator prioritizes maximizing the number of users served while still strictly adhering to their service requirements. This problem is defined formally as:
\begin{IEEEeqnarray}{cl}
\underset{
  \substack{
    \mathbf{x}, \mathbf{y}, \mathbf{w}
  }
}{\text{max}} \quad &
  \displaystyle\sum_{u\in\mathcal{U}} y_u(t) \\[0.5ex]
\textit{s.t.}\quad
& y_u(t)\, t_u\bigl(t, \mathbf{y}(t), \mathbf{x}(t), \mathbf{w}(t)\bigr) 
    \le T_u^{\max}, 
\nonumber\\*
& \hspace{12.6em} \forall u \in \mathcal{U} \label{c:p2_delay} \\[0.6ex]
& y_u(t)\, R_u^s\bigl(t, w_u(t)\bigr) \ge R_u^{\min},
    \quad \forall u \in \mathcal{U} \label{c:p2_throughput} \\[0.6ex]
& \sum_{g\in\mathcal{G}} x_u^g(t) = y_u(t),
    \quad \hspace{3.8em} \forall u \in \mathcal{U} \label{c:p2_association} \\[0.6ex]
& \sum_{u\in\mathcal{U}} w_u(t) \le B_s 
    \label{c:p2_service_bw} \\[0.6ex]
& \sum_{u\in\mathcal{U}} \sum_{g\in\mathcal{G}} x_u^g(t)\, C_g\bigl(u, t, w_u(t)\bigr) 
    \le C_{\text{Sat}}^{\text{Cap}} \label{c:p2_sat_cpu}
\end{IEEEeqnarray}

In this scenario, FlexSAN dynamically decides admission ($y_u(t)$), architecture ($x_u^g(t)$), and bandwidth allocation ($w_u(t)$), ensuring maximum possible service availability without compromising QoS for admitted users (constraints~\eqref{c:p2_delay}, \eqref{c:p2_throughput}). This adaptive admission control directly addresses the fundamental limitation of static SAN architectures, which are unable to scale gracefully under congestion.

\section{FlexSAN: Analysis and Design} \label{algorithm}

The FlexSAN orchestration problem formulated in Section~\ref{problem} is an NP-hard Mixed-Integer Non-Linear Program (MINLP), making it intractable to obtain the global optimum in real time under highly dynamic SAN environments. To address this challenge, we propose the Two-stage Adaptive Greedy Orchestration (TAGO) algorithm, which efficiently achieves high-quality approximate solutions with low computational complexity.

\subsection{TAGO: A Three-Tier Adaptive Framework}
The high-level structure of TAGO is presented in Algorithm~\ref{alg:tago}. It first assesses the network's congestion level and then routes the problem to the appropriate specialized algorithm: Cost-Efficient Orchestration (CEO) for $P_1$ or Service Maximization Orchestration (SMO) for $P_2$.

TAGO efficiently computes the congestion score, enabling rapid assessment of network conditions:
\begin{equation}
\sigma = \max \left( \frac{\sum_{u \in \mathcal{U}} C_{\text{min}}(u)}{C_{\text{Sat}}^{\text{Cap}}}, \frac{\sum_{u \in \mathcal{U}} w_u^{\text{min}}}{B_s} \right)
\label{eq:severity_score}
\end{equation}

where $w_u^{\text{min}}$ denotes the minimum bandwidth required for user $u$ to satisfy their throughput demand $R_u^{\text{min}}$, and $C_{\text{min}}(u)$ is the corresponding computational cost under the on-board gNB-DU architecture. $w_u^{\text{min}}$ for each user is efficiently computed using binary search with a tolerance of 0.01\,MHz. $\sigma$ represents the most optimistic resource utilization. Based on this score, TAGO employs a three-tier routing mechanism governed by thresholds $\sigma_{\text{light}}$ and $\sigma_{\text{heavy}}$:

As shown in Algorithm~\ref{alg:tago}, TAGO adopts different strategies according to the system load. Under light load conditions ($\sigma < \sigma_{\text{light}}$), TAGO prioritizes cost minimization by invoking the CEO algorithm. For marginal load ($\sigma_{\text{light}} \leq \sigma \leq \sigma_{\text{heavy}}$), it first attempts to apply CEO within a strict timeout $\tau_{\text{strict}}$ (e.g., 10\,ms), switching to the SMO algorithm if CEO fails to complete in time. When the system is heavily loaded ($\sigma > \sigma_{\text{heavy}}$), TAGO directly employs SMO to maximize user admissions.

\subsection{Cost-Efficient Orchestration (CEO) for $P_1$}

Under light-to-moderate load, CEO (Algorithm~\ref{alg:ceo}) aims to minimize OPEX while serving all users. It decouples the problem into two phases.

\subsubsection{\textbf{Phase 1}. Delay-Margin-Based Architecture Selection}

The key to minimizing cost is to assign as many users as possible to the cheaper on-board gNB-DU architecture. Our delay-margin-based strategy calculates the available slack in each user's delay budget to make this choice without compromising user service requirements. Define the effective delay margin for user $u$ under architecture $g$ as:
\begin{equation}
\Delta_u^g = T_u^{\text{max}} - t_u^{\text{fixed}}(g) - t_u^{\text{queue}}(\rho)
\end{equation}

where $t_u^{\text{fixed}}(1)$ represents all fixed delays for the gNB-DU, and $t_u^{\text{queue}}(\rho)$ represents the expected queuing delay under system utilization $\rho$. The architecture selection follows a cost-efficient greedy rule with margin threshold $\tau_{\text{margin}}$:
\begin{equation}
g_u^* = 
\begin{cases}
1~\text{ (gNB-DU)}, 
    & 
    \begin{array}[t]{@{}l@{}}
    \text{if } \Delta_u^1 > \tau_{\text{margin}}~\text{and} \\
    \hspace{-0.2em} C(1, w_u) < C(0, w_u)
    \end{array} \\[2ex]
0~\text{ (gNB)}, 
    & \text{otherwise}
\end{cases}
\end{equation}

A user is assigned the cost-efficient on-board gNB-DU ($g=1$) only if their latency can be comfortably met; otherwise, the low-latency on-board gNB ($g=0$) is chosen.
\begin{algorithm}[t]
\caption{Two-stage Adaptive Greedy Orchestration (TAGO)}
\label{alg:tago}
\begin{algorithmic}[1]
\Require User set $\mathcal{U}$, System capacities $(B_s, C_{\text{Sat}}^{\text{Cap}})$
\Ensure Admission $\mathbf{y}$, Architecture $\mathbf{x}$, Bandwidth $\mathbf{w}$
\State $\sigma \gets$ ComputeCongestionScore($\mathcal{U}$, $B_s$, $C_{\text{Sat}}^{\text{Cap}}$) \Comment{Eq.~\eqref{eq:severity_score}}
\If{$\sigma < \sigma_{\text{light}}$} \Comment{Light load: cost efficiency}
  \State $(\mathbf{y}, \mathbf{x}, \mathbf{w}) \gets$ CEO($\mathcal{U}$, $B_s$, $C_{\text{Sat}}^{\text{Cap}}$)
\ElsIf{$\sigma \in [\sigma_{\text{light}}, \sigma_{\text{heavy}}]$} \Comment{Marginal: CEO with 10ms timeout}
  \State $(\text{feasible}, \mathbf{x}', \mathbf{w}') \gets$ CEO($\mathcal{U}$, $B_s$, $C_{\text{Sat}}^{\text{Cap}}$, 10ms)
  \If{not feasible or partial admission}
      \State $(\mathbf{y}, \mathbf{x}, \mathbf{w}) \gets$ SMO($\mathcal{U}$, $B_s$, $C_{\text{Sat}}^{\text{Cap}}$)
  \Else
      \State $\mathbf{y} \gets \mathbf{1}$, $\mathbf{x} \gets \mathbf{x}'$, $\mathbf{w} \gets \mathbf{w}'$
  \EndIf
\Else \Comment{Heavy load: maximize admissions}
  \State $(\mathbf{y}, \mathbf{x}, \mathbf{w}) \gets$ SMO($\mathcal{U}$, $B_s$, $C_{\text{Sat}}^{\text{Cap}}$)
\EndIf
\State \Return $(\mathbf{y}, \mathbf{x}, \mathbf{w})$
\end{algorithmic}
\end{algorithm}

\subsubsection{\textbf{Phase 2}. Gradient-Based Bandwidth Optimization}

With the architectures fixed, we optimize bandwidth allocation using a gradient-based iterative refinement process that leverages the asymmetric computational cost structures of the two regenerative architectures. The computational cost of an on-board gNB user is far more sensitive to bandwidth changes than the on-board gNB-DU user. Our gradient-based approach leverages this key insight. When the initial allocation exceeds system capacity, we apply gradient-based compression:
\begin{equation}
w_u^{(k+1)} = w_u^{(k)} - \eta \cdot \nabla_w C(g_u, w_u^{(k)})
\label{eq:gradient_update}
\end{equation}

Here, $\eta$ denotes the bandwidth step size (e.g., 1 kHz, 2 kHz), and $\nabla_w C(g, w)$ represents the gradient of the computational cost with respect to the bandwidth. The algorithm prioritizes bandwidth reduction for users with higher cost gradients (gNB users have higher gradients than gNB-DU users), achieving efficient resource utilization while maintaining service demand constraints. The iterative refinement continues for up to $K_{\text{max}}$ iterations, adjusting bandwidth allocations for users with unmet service requirements and rebalancing from users with excess delay margins.

\begin{algorithm}[t!]
\caption{Cost-Efficient Orchestration (CEO)}
\label{alg:ceo}
\begin{algorithmic}[1]
\Require User set $\mathcal{U}$, Capacities $(B_s, C_{\text{Sat}}^{\text{Cap}})$, [Timeout $T_{\text{limit}}$]
\Ensure (feasible, $\mathbf{x}$, $\mathbf{w}$)
\State \textbf{Phase 1: Delay-Margin Architecture Selection}
\For{each user $u \in \mathcal{U}$}
  \State $\Delta_u^{\text{gNB-DU}} \gets T_u^{\text{max}} - t_u^{\text{fixed}}(\text{DU})$ \Comment{DU delay margin}
  \State Assign gNB-DU if $\Delta_u^{\text{gNB-DU}} > \tau_{\text{margin}}$, else gNB
\EndFor
\State \textbf{Phase 2: Bandwidth Optimization}
\State Initialize: $w_u \gets w_u^{\text{min}}$ for all $u \in \mathcal{U}$
\If{$\sum_{u} w_u > B_s$} \Comment{Apply gradient compression}
  \State Sort users by $\nabla_w C(g_u, w_u)$ and compress bandwidth
  \State Ensure all service demands remain satisfied
\EndIf
\State \textbf{Iterative Refinement}
\While{$\exists u: t_u > T_u^{\text{max}}$ and iterations $< K_{\text{max}}$}
  \State Increase bandwidth for the user with demand unmet
  \State Rebalance from users with excess margin if needed
\EndWhile
\State \Return (all service demands met, $\mathbf{x}$, $\mathbf{w}$)
\end{algorithmic}
\end{algorithm}

\subsection{Service Maximization Orchestration (SMO) for $P_2$}

Under heavy load conditions, SMO maximizes the number of admitted users by prioritizing users and packing resources.

\subsubsection{\textbf{Phase 1}. Composite-Score-Based Admission}
When resources are scarce, we must choose which users to admit. The optimal choice is not just the "cheapest" user. A user might be cheap but consume a bottleneck resource, or have an extremely strict service demand, making the system fragile. We introduce a composite efficiency score, $\eta_u$, to holistically evaluate a user's "admissibility".
\begin{equation}
\eta_u = \omega_{\text{eff}} \cdot \frac{1}{w_u^{\text{min}} + \alpha C_{\text{min}}(u)} + \omega_{\text{flex}} \cdot \frac{T_u^{\text{max}}}{\bar{T}}
\label{eq:composite_score}
\end{equation}

To support user admission across heterogeneous resources, a composite score $\eta_u$ is defined to jointly capture \textit{resource efficiency} and \textit{service demand flexibility}. The former favors users requiring fewer resources, while the latter prioritizes those with relaxed service constraints, facilitating easier accommodation. The score formulation incorporates a reference delay $\bar{T}$ under congestion and two non-negative weighting factors, $\omega_{\text{eff}}$ and $\omega_{\text{flex}}$, satisfying $\omega_{\text{eff}} + \omega_{\text{flex}} = 1$. A higher $\omega_{\text{eff}}$ emphasizes efficiency, whereas a higher $\omega_{\text{flex}}$ reflects flexibility. Additionally, a conversion factor $\alpha$ serves as the resource exchange rate, translating computational demand (e.g., GOPS) into equivalent bandwidth cost (e.g., Hz), thereby enabling unified evaluation across resource types. Users are ranked in descending order of $\eta_u$, and a greedy admission process sequentially accepts users with minimal resource configurations until either computational or bandwidth capacity is exhausted.

\subsubsection{\textbf{Phase 2}. Iterative Resource Compression}

The refinement phase employs an iterative compression strategy to maximize admissions. For each high-priority rejected user, the algorithm searches for admitted on-board gNB users who can switch to on-board gNB-DU architecture while still meeting their service requirements. If the freed resources (both computational and bandwidth) are sufficient to accommodate the rejected user, the architecture swap is performed, and the user is admitted. This process continues iteratively, prioritizing rejected users with higher composite scores.

\subsection{Complexity Analysis}
\begin{thm}\label{thm1}
The Congestion Assessment phase in the TAGO algorithm has a time complexity of $O(N)$, where $N$ denotes the number of users.
\end{thm}

\renewcommand{\IEEEQED}{\IEEEQEDopen}
\begin{IEEEproof}[Proof]
The Congestion Assessment phase computes the minimum required bandwidth for each user. As the bandwidth search space is bounded by the total system bandwidth (a constant), this operation scales linearly with the number of users. Thus, its complexity is $O(N)$.
\end{IEEEproof}

\begin{thm}\label{thm2}
The overall time complexity of the TAGO algorithm is $O(N \log N)$, dominated by sorting operations in both the CEO and SMO algorithms.
\end{thm}

\renewcommand{\IEEEQED}{\IEEEQEDopen}
\begin{IEEEproof}[Proof]
The dominant operation in both CEO and SMO is sorting, which has a time complexity of $O(N \log N)$. All other operations, including SMO's resource compression, are at most $O(N)$ and thus asymptotically insignificant. As TAGO executes an O(N) preprocessing step followed by either CEO or SMO, its overall complexity is $O(N \log N)$.
\end{IEEEproof}

\begin{algorithm}[t!]
\caption{Service Maximization Orchestration (SMO)}
\label{alg:smo}
\begin{algorithmic}[1]
\Require User set $\mathcal{U}$, Capacities $(B_s, C_{\text{Sat}}^{\text{Cap}})$
\Ensure ($\mathbf{y}$, $\mathbf{x}$, $\mathbf{w}$)
\State \textbf{Phase 1: Composite-Score-Based Greedy Admission}
\State Calculate composite score $\eta_u$ for each user \Comment{Eq.~\eqref{eq:composite_score}}
\State Sort users by $\eta_u$ in descending order
\State Greedily admit users with minimum configurations until resources exhausted
\State \textbf{Phase 2: Iterative Admission Refinement}
\For{high-priority rejected users}
  \State Find gNB users that can switch to gNB-DU and meet service demand
  \State If switching frees sufficient resources, perform swap and admit
\EndFor
\State \Return ($\mathbf{y}$, $\mathbf{x}$, $\mathbf{w}$)
\end{algorithmic}
\end{algorithm}

\section{Performance Evaluation} \label{performance}

In this section, we conduct comprehensive simulations to evaluate the performance of our proposed FlexSAN framework. Our goal is to:
\begin{itemize}
    \item The significant performance gains of FlexSAN orchestration compared to static SAN architectures.
    \item The effectiveness and efficiency of our proposed Two-Stage Adaptive Greedy Orchestration (TAGO) algorithm.
    \item The robustness of FlexSAN's adaptation to realistic, time-varying network traffic and service demands.
\end{itemize}

\subsection{Methodology and Experiments Setup}

\subsubsection{Simulation Environment and Parameters}

\begin{figure}[t!]
  \centering
  \includegraphics[width=0.95\columnwidth]{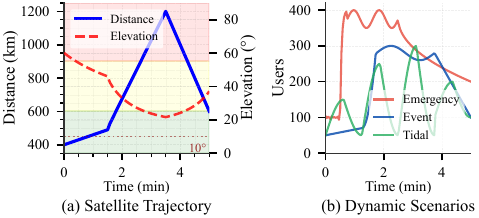}
  \caption{Satellite-Ground Distance and Dynamic User Traffic Patterns}
  \label{fig:setup}
\end{figure}

We simulate a low Earth orbit (LEO) satellite network featuring a single satellite with a reconfigurable payload. The simulation is based on the STARLINK-11671 [DTC] satellite's TLE data \cite{starlink_directtocell, celestrak_tle}, operating over a 20 MHz bandwidth in the 2 GHz S-band. The simulation accurately models key parameters, including orbital altitude, pass duration, and propagation delay. Ground users are uniformly distributed within a 25 km radius of a central ground station. The simulated time window is 5 minutes, during which the satellite's distance to the ground station varies as shown in Fig.~\ref{fig:setup}(a), with a simulation granularity of 100\,ms.

The satellite offers 32,000 GOPS of computational capacity, and resource allocation and computation costs for different functional splits follow the methodologies of \cite{10901285, 9107209}. Based on our OAI-based prototype profiling (see Section~\ref{motivation}), the additional F1 interface delay $t_{f1}$ for the on-board gNB-DU architecture is empirically set to $80$-$100$~ms. The data packet size $S_p$ is set to 1\,KB. The detailed parameters used in the TAGO algorithm are shown in Table~\ref{tab:tago-params}. These parameter values are empirically tuned to balance resource efficiency and service flexibility.


\begin{table}[t!]
  \centering
  \caption{Key Parameters of the TAGO Algorithm}
  \label{tab:tago-params}
  \begin{minipage}{0.9\linewidth}
    \renewcommand{\arraystretch}{1.15} 
    \setlength{\tabcolsep}{8pt} 
    \begin{tabular*}{\linewidth}{@{\hskip 12pt}c@{\extracolsep{\fill}}c@{\extracolsep{\fill}}c@{\hskip 12pt}}
      \toprule
      \textbf{Parameter Group}     & \textbf{Symbol(s)}              & \textbf{Value(s)} \\
      \midrule
      Congestion threshold         & $\sigma_{\text{light}},\ \sigma_{\text{heavy}}$     & $0.7,\ 1.0$ \\
      Timeout/Margin (ms)          & $\tau_{\text{strict}}$,\ $\tau_{\text{margin}}$    & $10,\ 5$ \\
      Bandwidth step (MHz)         & $\eta$                             & $0.05$ \\
      SMO eff./flex. weight        & $\omega_{\text{eff}},\ \omega_{\text{flex}}$ & $0.6,\ 0.4$ \\
      Bandwidth opt. iters         & ---                             & $5$--$12$ \\
      Admission refine iters       & ---                             & $\leq 3$ \\
      \bottomrule
    \end{tabular*}
  \end{minipage}
\end{table}

\subsubsection{User Service Requirements}

Each user requests a data rate ($R^{\text{min}}$) uniformly distributed between 0.5 and 2 Mbps, covering typical IoT to video streaming applications. To evaluate system performance under varied conditions, we define three service profiles based on maximum tolerable end-to-end delay ($T^{\max}$):

\begin{itemize}
\item STRICT: 70\% of users require $\le50$ ms delay, modeling critical services.
\item MIXED: Users are evenly distributed across delay tiers of $\le50$ ms, $\le100$ ms, and $\le200$ ms.
\item RELAXED: 70\% of users tolerate up to $\le200$ ms delay, simulating best-effort services.
\end{itemize}

Unless specified otherwise, we adopt the STRICT profile, representing the most stringent scenario to best illustrate architectural trade-offs.

\subsubsection{Traffic Scenarios}

To thoroughly evaluate FlexSAN, we simulate three representative traffic patterns, as illustrated in Fig.~\ref{fig:setup}(b).
\begin{itemize}
\item Emergency Response: A disaster scenario with four phases: (i) steady-state (100 users), (ii) rapid spike to 375 users within 30 seconds, (iii) sustained peak, and (iv) gradual recovery over 2.5 minutes, emulating urgent, large-scale connectivity needs.
\item Event-Driven: Represents event-related surges, including (i) pre-event linear growth (50-70 users), (ii) quadratic rise to 280 users at event onset, (iii) peak traffic with sinusoidal fluctuation ($\pm20$ users), and (iv) sharp post-event drop.
\item Tidal Traffic: Emulates diurnal patterns with four sinusoidal waves (amplitudes: 150, 250, 300, 200 users) over 5 minutes, reflecting time zone-based user migrations.
\end{itemize}

\begin{figure}[t!]
  \centering
  \includegraphics[width=\columnwidth]{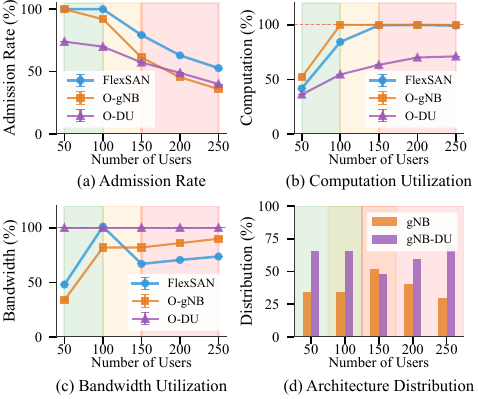} 
  \caption{Performance comparison of FlexSAN against static architectures.}
  \label{fig:exp1}
\end{figure}

\subsubsection{Baselines}

We compare FlexSAN against two baseline categories, tailored to distinct evaluation objectives.

\paragraph{SAN Architecture Baselines}
To highlight the value of architectural flexibility, we benchmark FlexSAN against two static schemes:
\begin{itemize}
\item On-board gNB (O-gNB): All users are served by a full on-board gNB, prioritizing low latency but limited by computational resources.
\item On-board gNB-DU (O-DU): All users are served by an on-board gNB-DU, with CU functions on the ground. This reduces on-board computation but cannot meet strict latency demands.
\end{itemize}


\paragraph{Algorithmic Baselines}
To evaluate our proposed TAGO algorithm, we benchmark against:
\begin{itemize}
\item Optimal: The theoretical upper bound, obtained by directly solving the Mixed-Integer Non-Linear Program (Section IV) with the Gurobi 12 solver\cite{gurobi}. While optimal, this approach is computationally impractical for real-time use.
\item Greedy: A heuristic that iteratively minimizes immediate resource cost, serving as a baseline to demonstrate the advantage of TAGO's two-stage decision logic.
\end{itemize}


\begin{figure}[t!]
  \centering
  \includegraphics[width=\columnwidth]{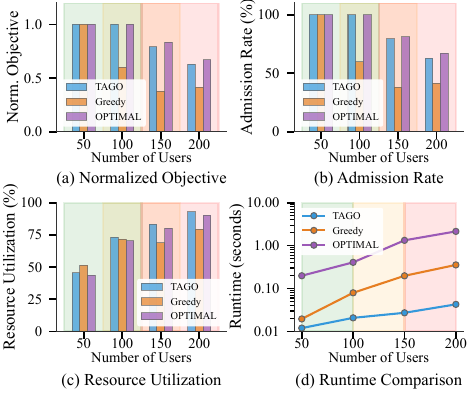} 
  \caption{Performance evaluation of the TAGO algorithm against baselines.}
  \label{fig:exp2}
\end{figure}

\subsection{Effectiveness of FlexSAN}

\subsubsection{Efficiency and Cost-Effectiveness under Light Load} 




Under light load conditions (e.g., 50 users), FlexSAN prioritizes minimizing operational expenditure (OPEX) while ensuring full service coverage. As shown in Fig.~\ref{fig:exp1}(a), both FlexSAN and the on-board gNB architecture achieve 100\% user admission. In contrast, the on-board gNB-DU architecture admits only 74\% of users, failing to meet latency constraints despite having sufficient resources.

Although the on-board gNB guarantees full admission, it incurs higher operational costs by consuming 52.0\% of onboard computational resources. FlexSAN, by comparison, achieves the same admission rate with only 41.8\% computational usage, translating to an OPEX reduction of nearly 15\% (see Fig.\ref{fig:exp1}(b)). This efficiency is enabled by FlexSAN's intelligent orchestration, which dynamically assigns 66\% of users with relaxed QoS requirements to the more cost-effective gNB-DU, while allocating only 34\% of latency-sensitive users to the on-board gNB (Fig.\ref{fig:exp1}(d)).

As a result, FlexSAN slightly increases bandwidth consumption (48.0\%) but conserves valuable computational resources. In contrast, the on-board gNB-DU architecture fully saturates bandwidth (100\%) yet still fails to satisfy latency demands (Fig.~\ref{fig:exp1}(c)), demonstrating clear inefficiencies. These findings highlight FlexSAN's ability to balance universal QoS guarantees with operational cost-effectiveness under light traffic conditions.

\subsubsection{Admission Maximization under Heavy Load} 
As user load increases, FlexSAN dynamically adapts its resource management strategy to maximize user admission, outperforming static architectures that quickly encounter bottlenecks. For instance, the on-board gNB saturates computational resources at only 100 users (Fig.~\ref{fig:exp1}(b)), while the on-board gNB-DU continues to struggle due to latency and bandwidth constraints.

Under heavy load conditions (150-250 users), FlexSAN consistently achieves 29\%-47\% higher admission rates compared to the best-performing static architecture. Specifically, at 250 users, FlexSAN admits 52.8\% of users, significantly outperforming the on-board gNB (36\%) and gNB-DU (40\%).

This performance gain stems from FlexSAN's flexible orchestration. When computational capacity becomes the dominant bottleneck, FlexSAN allocates only 29.5\% of users-those with the most stringent latency demands-to the on-board gNB, and offloads the remaining 70.5\% to the gNB-DU (Fig.~\ref{fig:exp1}(d)). This allocation improves bandwidth utilization and relieves computational pressure. As a result, FlexSAN achieves near-optimal resource utilization, with both computational and bandwidth usage reaching up to 99.6\%, effectively overcoming early saturation and enhancing overall system capacity.



\subsection{Effectiveness of the TAGO Algorithm}
Our proposed TAGO algorithm is designed to deliver near-optimal solutions with minimal computational overhead, making it practical for real-time deployment. Fig.~\ref{fig:exp2} validates its dual superiority in both solution quality and efficiency.

\subsubsection{Near-Optimal Solution Quality}
As shown in Fig.~\ref{fig:exp2}(a), TAGO achieves normalized objective values within 5.5\% of the MINLP optimum across all user scales. This high fidelity results in user admission rates averaging within 6\% of the MINLP-optimal solution (Fig.~\ref{fig:exp2}(b)), substantially outperforming the Greedy heuristic, which can lag by up to 26.2\%. Furthermore, TAGO's resource utilization closely matches the optimal benchmark (Fig.~\ref{fig:exp2}(c)), confirming its efficiency.

\begin{figure}[t!]
  \centering
  \includegraphics[width=0.9\columnwidth]{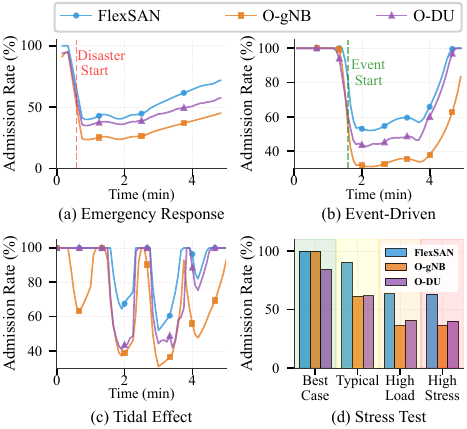} 
  \caption{Performance of FlexSAN in dynamic real-world traffic scenarios.}
  \label{fig:exp3}
\end{figure}

\subsubsection{Real-Time Computational Efficiency} 
The most compelling advantage of TAGO is its computational speed. Fig.~\ref{fig:exp2}(d), plotted on a logarithmic scale, starkly illustrates the difference in runtime. At a scale of 200 users, the MINLP solver requires nearly 12 seconds, rendering it unsuitable for dynamic environments. In contrast, TAGO finds a near-optimal solution in just 67 milliseconds, achieving a speedup of more than 170 times. This adherence to polynomial time complexity, as opposed to the exponential growth of the MINLP solver, confirms that TAGO is a practical and pragmatic algorithm for real-time orchestration in LEO satellite networks.

\subsection{Robustness of FlexSAN}
\subsubsection{Adaptability in Realistic Dynamic Scenarios}
Beyond static loads, we evaluated FlexSAN's ability to adapt to realistic, time-varying traffic patterns, where its superiority is even more evident. Fig.~\ref{fig:exp3} showcases its resilience and adaptability across several challenging scenarios.

In the Emergency Response scenario (Fig.~\ref{fig:exp3}(a)), a sudden surge in demand causes the performance of static architectures to collapse to below 30\% admission. FlexSAN, however, demonstrates graceful degradation, maintaining an admission rate above 40\% during the initial shock and recovering swiftly to over 70\%, effectively serving more than twice as many users as static systems during the critical recovery phase.

Similarly, in the Event-Driven scenario (Fig.~\ref{fig:exp3}(b)), FlexSAN effectively handles the sharp traffic spike, maintaining an admission rate above 50\% at the peak, while static methods falter. In the Tidal Effect scenario (Fig.~\ref{fig:exp3}(c)), FlexSAN's performance curve is significantly smoother and consistently higher than the volatile and inefficient performance of the static baselines, indicating a more stable and reliable service for users.

\subsubsection{Robustness Under Multi-Dimensional Stress}
Finally, the multi-dimensional Stress Test (Fig.~\ref{fig:exp3}(d)) confirms FlexSAN's robustness. Under the most demanding "High Stress" conditions (high load, strict service requirements, and long propagation delays), FlexSAN achieves a 63.5\% admission rate, nearly doubling the 37.8\% rate of the best-performing static alternative. These results collectively prove that FlexSAN is not only efficient but also highly adaptive and resilient, making it a robust solution for the unpredictable nature of next-generation satellite communications.

\vspace{4pt}
\section{Conclusions} \label{conclusion}
Balancing OPEX and user QoS is a key challenge in SANs due to the rigidity of static regenerative architectures. This paper presents FlexSAN, a dynamic regenerative SAN architecture employing a two-stage optimization strategy: minimizing computational cost under normal load and maximizing user admission during congestion. To address the NP-hardness of the problem, a heuristic algorithm, TAGO, is developed to enable flexible user admission, adaptive payload selection, and dynamic bandwidth allocation. Experimental results show that FlexSAN improves user admission rates by 36.1\% and reduces OPEX by 15\% compared to static SANs.

  
\clearpage
\bibliographystyle{IEEEtran}
\bibliography{main}
\end{document}